\documentclass[slac_one]{revtex4}
\usepackage{graphicx}
\usepackage{fancyhdr}
\newcommand{\me}{\mbox{${\rm \not\! E}$}}
\newcommand{\met}{\mbox{${\rm \not\! E}_{\rm T}$}}

\begin{document}

\title{{\small{Hadron Collider Physics Symposium (HCP2008),
Galena, Illinois, USA}}\\ 
\vspace{12pt}
Higgs Boson Beyond the SM (Alternate EW Breaking Scenarios): Nonstandard Higgs Decays} 

%

\author{S. Chang}
\affiliation{Center for Cosmology and Particle Physics, Dept. of Physics, New York University, New York, NY 10003}

\begin{abstract}
The crucial search for the Higgs boson at future colliders is capable of discovering the Standard Model Higgs, but is not  guaranteed to discover a Higgs that decays nonstandardly.  Such new physics is motivated by many aspects; from experiment, by the tension between precision electroweak fits and the LEP2 direct search bound and from theory, by issues of satisfying the LEP2 bound in theories that naturally have a lighter Higgs.  The consistent nonstandard Higgs scenario is that the Higgs is lighter than the LEP2 bound and is consistent with direct search limits because the dominant new decays are cascades into a multi-body final state, mediated by new light particles.  The reduction in the Standard Model branching ratio implies that searches dependent on these decays can be severely weakened.  Thus, it is imperative to design searches capable of discovering the new decays.  Some of the possible analyses that can be performed at the Tevatron and LHC are presented.  However, much work remains to see if discovery of the nonstandard Higgs at hadron colliders can be ensured.
\end{abstract}

\maketitle

\thispagestyle{fancy}


\section{INTRODUCTION} 
The current searches for the Higgs bosons at Tevatron Run II and future ones at the LHC are picking up where LEP2 left off.  The Higgs boson of the Standard Model (SM) has yet to be discovered and has a lower bound of 114.4 GeV on its mass from the LEP2 SM search \cite{Barate:2003sz}.  The importance of discovering the Higgs cannot be overemphasized, as it is a crucial discriminant between different models of electroweak symmetry breaking.  In particular, from notions of naturalness from the hierarchy problem, it is expected to be accompanied by new physics, the best known example being supersymmetry.  Reflecting the importance of this discovery, the ATLAS and CMS experiments have dedicated their detectors to be capable of discovering the SM Higgs if it is lighter than 1 TeV.  For instance, the recent analyses by CMS demonstrate that given 30 fb$^{-1}$ of luminosity, the Higgs can be discovered up to $\sim 700$ GeV \cite{Ball:2007zza}.

However, a potential loophole to this reach is if the Higgs has nonstandard properties.  In particular, there has been a lot of recent work on nonstandard Higgs decays, where the Higgs boson production is unchanged, but the dominant decay is no longer SM-like and is instead into a multi-body final state.  For a recent review, see \cite{Chang:2008cw}.  With these new decays, the mass of the Higgs can become lighter than the LEP2 bound, while unfortunately making future Higgs searches based on standard decays less effective.  

The rest of the talk is as follows:  in section \ref{sec:motivation}, we will discuss the hints from both experimental and theoretical aspects that suggest nonstandard Higgs decays;  in section \ref{sec:examples}, we will provide example scenarios and discuss their phenomenology; in section \ref{sec:tevatron} and \ref{sec:lhc}, we will discuss some analyses at respectively the Tevatron and LHC which can help probe these nonstandard decays; and finally in section \ref{sec:conclusions}, we conclude and provide some future outlook.         

\section{MOTIVATION FOR NONSTANDARD HIGGS DECAYS \label{sec:motivation}}
The motivations for nonstandard Higgs decays comes from several different directions \cite{Chang:2008cw}.  Each of these on their own may only be described as a hint, but taken together they make a compelling case.  Nonstandard Higgs decays provide a scenario in which all of these motivations can be made consistent.  However, aside from these considerations, the phenomenological implication that the Higgs could be missed, arguably motivates their exploration all on its own.  

In the next few subsections, it will be argued that the consistent scenario contains a Higgs lighter than the LEP2 bound.  Consistency with the bound is obtained due to new decays for the Higgs that dominate over the standard ones.  These decays are cascade decays into a multi-body final state, which are mediated by new particles lighter than the Higgs.  The Higgs production cross sections are essentially unchanged and thus, the phenomenology at colliders is completely specified by the Higgs decay topology.   

\subsection{EXPERIMENTAL}
Experimentally, there has been a long standing tension between the SM direct search limit from LEP2 and the indirect bounds from fits to precision electroweak observables (PEWOs).  In particular, PEWOs from LEP1, SLD and other experiments, have preferred a Higgs lighter than the LEP2 bound.  It has been emphasized that this preference is increased if the most discrepant observable, the forward-backward asymmetry of the $b$ quark $A^{(0,b)}_{FB}$ is removed from the fit \cite{Chanowitz:2002cd}.  This trend for a light Higgs has been reinforced by the new measurements of the top \cite{Heintz:2008ue} and $W$ \cite{Trischuk:2008zz} mass from Tevatron Run-II .  The current fit, including all data, puts an upper bound of 160 GeV at 95\% CL with a central value of 87 GeV \cite{lepewwg}.  Roughly this is at tension with the direct search limit at $1\sigma$.  

The nonstandard Higgs scenario can resolve this slight tension.  To do so, it requires the Higgs mass be lighter, around 90-100 GeV, and that it couples to SM particles with SM strength, as suggested by the PEWO fit.  It resolves the conflict with the direct search bound by having new decays for the Higgs which dominate over the standard decays, reducing the rate of SM decays.  The experimental limits on the new decays will be discussed in section \ref{sec:examples}.        
    
\subsection{THEORETICAL}
There are two aspects to the theoretical motivations for nonstandard Higgs decays.  The first is the issue of fine-tuning in beyond the Standard Model theories with an elementary Higgs boson.  There are many such theories where the Higgs boson is at tree level, lighter than the $Z$ boson.  This happens within the minimal supersymmetric Standard Model (MSSM) and also in many models where the Higgs is a pseudo-Nambu-Goldstone boson.  For a large part of the parameter space of these theories, the LEP2 bound applies to this Higgs boson, which is in tension with its tendency to be lighter.  To increase the mass of the Higgs requires increasing the mass of top quark partners which are responsible for canceling the quadratic divergences to the Higgs mass parameter as motivated by the hierarchy problem.  If these masses have to be pushed up to the TeV scale, proper electroweak symmetry breaking requires a tuning of about a few percent, introducing a residual fine-tuning issue often referred to as the ``little hierarchy" problem.  This fine-tuning is exponentially sensitive to the direct search bound on the Higgs mass.  Since the nonstandard Higgs can weaken the LEP2 bound, the mass can be brought closer to the natural values near the $Z$ mass.  Such motivations for supersymmetric models are discussed in \cite{Dermisek:2005ar}.    

The second theoretical aspect is that the SM Higgs decay width is very small for Higgses of mass around 100 GeV, making it susceptible to new decays.    Since the Higgs couples to particles proportional to their mass, plus the fact that the $W, Z,$ and top pair decays are not accessible for the Higgs, means that the decay width is small.  For instance, for a 100 GeV Higgs, $\Gamma_{SM}/{m_h} \sim 10^{-5}$.  Thus, in this mass range, it is highly susceptible to new decays dominating over the standard ones.  If there is a new particle, that is lighter than the Higgs mass, the Higgs could decay onshell into it and thus only a relatively weak coupling to the Higgs is required to make this new decay the dominant one.  Such new decays reduce the branching ratio to SM decays,  
\begin{equation}
BR(h \to SM) = \frac{\Gamma_{SM}}{\Gamma_{SM}+\Gamma_{new}}.
\label{eq:brreduction}
\end{equation}
Such light new states can appear in many beyond the Standard Model theories, and can be light as long as they are neutral and don't couple strongly to the $Z$ boson.  If this new particle is stable, the Higgs decay would be effectively invisible, but are strongly constrained \cite{:2001xz}.  Thus, these particles should be unstable and provide visible decay products.  This means that the decay is a cascade into a multi-body final state, facilitated by these new particles.  Since the LEP2 searches for the Higgs has strongly constrained most SM two-body decays \cite{Chang:2008cw}, these cascades naturally produce multi-body decays that were not specifically analyzed by LEP2, which can have weaker constraints.    

\subsection{PHENOMENOLOGY}
The LEP2 SM Higgs search constrains the branching ratio into the SM decay as a function of the Higgs mass \cite{Barate:2003sz}.  The limit for a 90-100 GeV Higgs, requires the SM branching ratio to be less than $25\%$.  Since most standard Higgs searches are statistics limited, the amount of luminosity required to discover a reduced branching ratio is
\begin{equation}
{\cal L} = \frac{{\cal L}_{old}}{BR(h\to SM)^2}.
\end{equation}   
For the maximal 25\% branching ratio allowed by LEP2 for a nonstandard Higgs in this mass range, a search that required 30 fb$^{-1}$ now requires $480$ fb$^{-1}$.  Extrapolating the CMS search plot, this would require $\gtrsim 250$ fb$^{-1}$ to discover a 100 GeV nonstandard Higgs.  Such luminosities require a few years of design luminosity running at the LHC.  This naive scaling doesn't take into account any losses of efficiency, triggering, or increases in background rates that moving to design luminosity can incur, but also doesn't take into account any gains in analysis power.  

At any rate, it is possible to severely weaken the Higgs searches that focus on standard decay topologies.  Thus, it is imperative to explore whether the nonstandard Higgs decays can themselves be discovered at the Tevatron and LHC.  These decays have many challenging issues at these colliders.  Due to the multi-body final state, the decay products are typically softer since they are sharing the energy from the Higgs mass, leading to issues with triggering and/or reconstruction.  However, since the decay topologies are generic, there is a possibility that these Higgs decays can be picked up by searches for other signal hypotheses.  The fact that the Higgs production mechanisms are unchanged allows use of any associated objects in the production, for instance, forward tagging jets from vector boson fusion or $W, Z$ boson in associated production.  Also, in the general class of possibilities, there could be highly displaced vertices from the decay of the facilitating intermediate particle.            

\section{EXAMPLE SCENARIOS \label{sec:examples}}
Nonstandard Higgs decays can occur in many models containing a Higgs boson. What is required is a new particle which is allowed to be lighter than the Higgs, that can couple strong enough to the Higgs to dominate over the standard Higgs decays, and that is also unstable.  Most of the model building has been on the supersymmetric side (for references, see \cite{Chang:2008cw}), but it has also been explored in models with hidden valleys \cite{Strassler:2006ri} and new neutrino physics \cite{Graesser:2007yj,Graesser:2007pc,de Gouvea:2007uz,Chang:2007de}.  Instead of going through the particulars of the models, we will briefly summarize the phenomenology by just focusing on the types of new particles and nonstandard Higgs decay topologies they generate.  These scenarios are summarized in Table \ref{tab:examples}.    

\begin{table}[t]
\begin{center}
\caption{Example Nonstandard Higgs Scenarios \\}
\begin{tabular}{|c|c|c|c|}
\hline Higgs Decay & Subsequent Decay & Full Higgs Topology & Mass limit (GeV)\\
\hline 
\hline
$h\to 2X$ & $X\to b\bar{b},\tau \bar{\tau}$& $4b, 2b\, 2\tau, 4\tau \cdots$& 4$b$: 110, $4\tau$: 86 \cite{Sopczak:2006vn}\\
\hline
$h\to 2X$ & $X\to 2g, 2\gamma$& $4g, 2g\, 2\gamma, 4\gamma$ & 4$g$: $90-100$  \cite{Chang:2005ht}\\
\hline
$h\to 2X$ & $X\to 3q$ &$6q$ & $ 82$ \cite{Carpenter:2007zz}\\
\hline
$h\to X_1 X_2$ & $X_2 \to X_1 \, f\bar{f}$ & $f\bar{f} + \me$ & $100$ \cite{Chang:2007de}\\
\hline
\end{tabular}
\label{tab:examples}
\end{center}
\end{table}

\subsection{ONE NEW PARTICLE}
The simplest example is where a single new particle $X$ is introduced.  In this case, the nonstandard decay is $h\to 2X$, followed by the $X$ decays.  There are a few well motivated possibilities for how $X$ decays.  If $X$ is a scalar or pseudo-scalar, it can pick up Yukawa couplings to SM fermions by mixing with any of the Higgs bosons of the theory; for e.g., in the MSSM, it can mix with the $h^0,H^0,$ or $A^0$.  Thus, like the Higgs, it decays preferentially  into heavy SM fermions.  Therefore, the final Higgs decay topology is into four fermions, predominantly $4b, 2b\, 2\tau,$ and $4\tau$ depending on if $X$'s mass is above the $2b$ or $2\tau$ threshold.  In this case, the limits from the LEP analysis of \cite{Sopczak:2006vn}, limit the Higgs mass to be heavier than 110 GeV if the decay is into 4$b$ or 86 GeV if it is into $4\tau$.  The weakness of the $4\tau$ limit illustrate the possibility of significantly weaker limits for some nonstandard decays.          

For scalar $X$'s, another possibility is that it is fermiophobic to decays into SM fermions.  If so, it can receive loop-induced decays, through heavy new particles in the loop, into light gauge bosons \cite{Dobrescu:2000jt, Chang:2005ht}.  Depending on its couplings to the new heavy particles, the decay can be into two gluons ($2g$) or two photons ($2\gamma$), giving the potential topologies of $4g, 2g\, 2\gamma,$  or $4\gamma$.  If the heavy particles are strongly interacting, the $h\to 4g$ decay dominates, which has no dedicated LEP2 analysis, but can be estimated to have a lower limit of about 90-100 GeV \cite{Chang:2005ht}.  

Finally, for fermionic $X$'s, there can be decays into three quarks, $X\to 3q$.  Such Higgs decays into $6q$ have weak limits and have been argued to be about 82 GeV \cite{Carpenter:2007zz}.  As mentioned in that paper, due to constraints on the R-parity violating couplings required, $X$'s decays can be highly displaced, giving two displaced vertices per Higgs decay.  The potential for displaced vertices was also discussed in \cite{Chang:2005ht,Strassler:2006ri}.  The capability to detect this distinguishing signature at LHCb will be discussed later in section \ref{sec:lhc}.      

\subsection{TWO NEW PARTICLES}
If there are two new particles $X_1$ and $X_2$, there are additional possibilities.  In this talk, we will focus on one particular scenario where the lightest one $X_1$ is stable and neutral and the heavier particle $X_2$ decays as $X_1 f\bar{f}$, where $f$ is some SM fermion \cite{Chang:2007de}.  Thus, the Higgs decay $h \to X_2 X_1 \to 2X_1 f\bar{f}$ contains both visible energy from the SM fermions and missing energy from the $X_1$'s.  Such decays have limits from LEP2 searches for $h\to WW^*$ and sparticle pair production.  These limits are strong, but depend on the assumed branching ratios for $X_2$ into the different SM fermions.  For the most part, the constraints in order of decreasing strength are if the fermions are neutrinos, charged leptons, $b$ quarks, and light quarks \cite{Chang:2007de}.  Depending on the scenario, the mass of the Higgs can be as light as 98 GeV.    

Other scenarios of this class, but that we won't have time to discuss, involve new neutrino physics, see  \cite{Graesser:2007yj,Graesser:2007pc,de Gouvea:2007uz,Chang:2007de}.   

\section{TEVATRON ANALYSES \label{sec:tevatron}}
The Tevatron analyses which can probe this scenario are best suited towards optimizing towards specific decays.  The first goal should be to push the SM Higgs search from both the mass range near the $WW$ threshold and if possible, that near the LEP2 bound.  The latter is the most important because improving upon the LEP2 bound exacerbates exponentially the fine-tuning issues from the ``little hierarchy" as mentioned in section \ref{sec:motivation}.  In the natural parameter space of theories beyond the SM, the only possibility will be for nonstandard Higgs decays to be occurring, giving clues to the Higgs physics required of these models.    

Tevatron searches can also look for particularly clean nonstandard Higgs topologies.  The first case in Table \ref{tab:examples}, where $X$ is light so that it dominantly decays into muons, produces a $4\mu$ decay \cite{Zhu:2006zv} which is being actively studied at D$\emptyset$ \cite{andyhaas}.  In particular, this probes the HyperCP interpretation of a light pseudoscalar $X$ of mass 214.3 MeV, decaying into two muons \cite{Park:2005eka}.  There is a good efficiency to reconstruct muons and the background is quite small, but unfortunately due to rates, this search will only be able to probe $X$ masses $<$ 450 MeV, where the $2\mu$ decay dominates \cite{andyhaas}.    In particular, the more natural possibility of $X$ being heavy enough to decay into $2\tau$ looks to be difficult at the Tevatron \cite{Graham:2006tr} and needs more good ideas.  There has also been some work on looking for the more constrained scenario of $4b$ decays \cite{Stelzer:2006sp,Cheung:2007sva,Carena:2007jk}.

\section{LHC ANALYSES \label{sec:lhc}}
LHC analyses have more freedom in being model-independent than Tevatron searches, due to advantages in Higgs production cross sections and ability to get integrated luminosity.  There are many analyses which have focused on searches that require design amounts of integrated luminosity, $>$ 100 fb$^{-1}$.  For instance, decays into $2g\, 2\gamma$ and $4\gamma$ were analyzed by \cite{Martin:2007dx, Chang:2006bw} and the search for $4\tau$ was explored in \cite{Forshaw:2007ra}.  However, in this talk, we will focus on analyses that will be doable in early running at the LHC.  

\subsection{VECTOR BOSON FUSION: $h\to l^+ l^- + \met$}
As an illustration of the robustness of some searches to have good efficiency for nonstandard Higgs decays, we have looked into an ATLAS study of a vector boson fusion produced Higgs that decays into $WW^*$, in the dilepton plus missing energy channel \cite{Asai:2004ws}.  Such searches can utilize the distinctive feature of vector boson fusion, namely two forward tagging jets and lack of hadronic activity between them.  In ongoing work \cite{thomas}, the selection cuts used in the ATLAS paper were studied for the decay $h\to X_2 X_1 \to l^+ l^- + \met$.  The signal events were generated in PYTHIA and PGS4 was used as a detector simulator.  Signal efficiencies as good as 2\% were found for the dielectron mode.  The ATLAS background estimate in this mode was 1.33 fb \cite{Asai:2004ws}.  Given that vector boson fusion production of a 100 GeV Higgs boson is around 4 pb and 30 fb$^{-1}$ of integrate luminosity, suggests that branching ratios $X_2 \to e\bar{e} X_1$ of 2\% can be discovered at 5$\sigma$.  This is just under the expected value for $X_2$ decays mediated by an offshell $Z$ boson.        

This study illustrates that nonstandard Higgs decays can be picked up by searches for a different signal hypothesis; in this case, a different Higgs decay.  This robustness should be preserved, if at all possible, in LHC searches for the Higgs.  This is because once studies become too signal specific, by using advanced techniques like neural networks, they lose their capability to pick up alternative signals.  Additionally, it is worth mentioning that excesses inconsistent with the intended signal should still be taken seriously in case they actually are new physics signals.   

\subsection{LHCb}
One nonstandard Higgs decay possibility, that of having highly displaced vertices, is extremely interesting and could be discovered at an unlikely LHC experiment.  LHCb with it's more lenient triggering and precise vertexing has an advantage to detect such signals.  Such a possibility has been studied by \cite{Kaplan:2007ap} and also by the LHCb collaboration \cite{Gueissaz:2007ds}, for the case of $h\to 6q$ (see Table \ref{tab:examples}) where there are two displaced vertices with three quarks each pointing back to them.   

LHCb's non-hermetic nature requires the Higgs to have a nontrivial boost in order for both vertices to appear in the detector.  These studies determined that a reasonable percentage of Higgs decays ($\sim$ few \%) are in this geometrical acceptance.  Another distinguishing feature is the enhanced charged track multiplicity at the displaced vertices.  Such double displaced vertex events have a small background \cite{Kaplan:2007ap}, the background is negligible.  Thus, this could be be discovered at LHCb in its early running phase.  Furthermore, rough mass information for the Higgs and $X$ can be extracted from the decay \cite{Gueissaz:2007ds}.      

\section{CONCLUSIONS AND OUTLOOK \label{sec:conclusions}}
In this talk, the motivation and phenomenology for nonstandard Higgs decays has been argued to be a crucial issue for future Higgs searches.  Such decays are a consistent scenario which resolve issues in the consistency between precision electroweak fits, direct search bounds on the Higgs, and fine-tuning issues within theories of beyond the Standard Model physics.  Its main phenomenological implication is that the standard decay is reduced in rate, making it more difficult and perhaps impossible to discover the Higgs in standard channels at the LHC.  This motivates looking at the phenomenology of the nonstandard decays themselves and determining if they can be discovered.  

Two simple examples were presented, the first with one new particle, where the Higgs decays into a completely visible decay containing $>= 4$ SM particles and the second with two new particles, where the Higgs decays into two SM fermions and missing energy.  Limits on these nonstandard decays can be quite weak and allow the Higgs mass to be around 100 GeV.

There are many analyses that can be used to probe this scenario at Tevatron and LHC.  Both experiments should look for the SM Higgs and if there is no signal, put stronger limits on a Higgs branching ratio into the SM decay.  Improving the LEP2 bound pushes the fine-tuning issue of the ``little hierarchy" problem and would motivate nonstandard Higgs decays if other signals of these beyond the SM theories are discovered.  Searches for the nonstandard Higgs decays themselves should be approached differently by Tevatron and LHC experiments.  The Tevatron should focus on model-dependent analyses, which try to maximize their analysis power.  An example at D$\emptyset$ was presented, a search for $h\to 4\mu$.  The LHC given its advantages in overall rates and statistics should strive to be model-independent, so as to cast a wide net which is capable of discovering many possibilities.  For instance, the ATLAS vector boson fusion analysis of $h\to WW^*$ has a good efficiency for discovering $h\to X_2 X_1 \to l^+ l^- + \met$.  In addition, if the Higgs decays with two highly displaced vertices, it could be discovered at LHCb.

The myriad possibilities of searching for such a Higgs demonstrate how nonstandard Higgs phenomenology can manifest itself in several unexpected possibilities.  If no SM Higgs signal is discovered after 30 fb$^{-1}$ is analyzed at LHC, it will be important to consider nonstandard Higgs scenarios as the next viable scenario for a Higgs boson to exist, but remain undiscovered.  That time period will be a good point in which to reevaluate what the focus of LHC Higgs analyses should be in the design luminosity phase.  Hopefully, given sufficient planning, it will be possible to determine an approach which will ensure the discovery of the Higgs boson.     

\begin{acknowledgments}
The author wishes to thank the organizers of the HCP2008 symposium for the talk invitation and for a well run and enjoyable conference. This work was supported by NSF CAREER grant PHY-0449818 and DOE grant \# DE-FG02-06ER41417.
\end{acknowledgments}


\begin{thebibliography}{99} 

\bibitem{Barate:2003sz}
  R.~Barate {\it et al.}  [LEP Working Group for Higgs boson searches and
                  ALEPH Collaboration and  and],
  Phys.\ Lett.\  B {\bf 565}, 61 (2003)
  [arXiv:hep-ex/0306033].

\bibitem{Ball:2007zza}
  G.~L.~Bayatian {\it et al.}  [CMS Collaboration],
  J.\ Phys.\ G {\bf 34}, 995 (2007).

\bibitem{Chang:2008cw}
  S.~Chang, R.~Dermisek, J.~F.~Gunion and N.~Weiner,
  arXiv:0801.4554 [hep-ph].

\bibitem{Chanowitz:2002cd}
  M.~S.~Chanowitz,
  Phys.\ Rev.\  D {\bf 66}, 073002 (2002)
  [arXiv:hep-ph/0207123].

\bibitem{Heintz:2008ue}
  U.~Heintz  [CDF - Run II and D0 - Run II Collaborations],
  arXiv:0806.1202 [hep-ex].

\bibitem{Trischuk:2008zz}
  W.~Trischuk,
  Nucl.\ Phys.\ Proc.\ Suppl.\  {\bf 177-178}, 22 (2008).


\bibitem{lepewwg}
  ``LEP-Electroweak Working Group",
	http://lepewwg.web.cern.ch/LEPEWWG/.


\bibitem{Dermisek:2005ar}
  R.~Dermisek and J.~F.~Gunion,
  Phys.\ Rev.\ Lett.\  {\bf 95}, 041801 (2005)
  [arXiv:hep-ph/0502105].

\bibitem{:2001xz}
    [LEP Higgs Working for Higgs boson searches, ALEPH, DELPHI, L3 and OPAL Collaborations],
  arXiv:hep-ex/0107032.

\bibitem{Strassler:2006ri}
  M.~J.~Strassler and K.~M.~Zurek,
  Phys.\ Lett.\  B {\bf 661}, 263 (2008)
  [arXiv:hep-ph/0605193].

\bibitem{Graesser:2007yj}
  M.~L.~Graesser,
  Phys.\ Rev.\  D {\bf 76}, 075006 (2007)
  [arXiv:0704.0438 [hep-ph]].

\bibitem{Graesser:2007pc}
  M.~L.~Graesser,
  arXiv:0705.2190 [hep-ph].
  
\bibitem{de Gouvea:2007uz}
  A.~de Gouvea,
  arXiv:0706.1732 [hep-ph].

\bibitem{Chang:2007de}
  S.~Chang and N.~Weiner,
  JHEP {\bf 0805}, 074 (2008)
  [arXiv:0710.4591 [hep-ph]].

\bibitem{Sopczak:2006vn}
  A.~Sopczak  [ALEPH, DELPHI, L3 and OPAL Collaborations],
  arXiv:hep-ph/0602136.
  
\bibitem{Chang:2005ht}
  S.~Chang, P.~J.~Fox and N.~Weiner,
  JHEP {\bf 0608}, 068 (2006)
  [arXiv:hep-ph/0511250].

\bibitem{Carpenter:2007zz}
  L.~M.~Carpenter, D.~E.~Kaplan and E.~J.~Rhee,
  Phys.\ Rev.\ Lett.\  {\bf 99}, 211801 (2007)
  [arXiv:hep-ph/0607204].


\bibitem{Dobrescu:2000jt}
  B.~A.~Dobrescu, G.~L.~Landsberg and K.~T.~Matchev,
  Phys.\ Rev.\  D {\bf 63}, 075003 (2001)
  [arXiv:hep-ph/0005308].

\bibitem{Zhu:2006zv}
  S.~h.~Zhu,
  arXiv:hep-ph/0611270.


\bibitem{andyhaas}
 Private conversation with Andy Haas.
 
\bibitem{Park:2005eka}
  H.~Park {\it et al.}  [HyperCP Collaboration],
  Phys.\ Rev.\ Lett.\  {\bf 94}, 021801 (2005)
  [arXiv:hep-ex/0501014].

\bibitem{Graham:2006tr}
  P.~W.~Graham, A.~Pierce and J.~G.~Wacker,
  arXiv:hep-ph/0605162.

\bibitem{Stelzer:2006sp}
  T.~Stelzer, S.~Wiesenfeldt and S.~Willenbrock,
  Phys.\ Rev.\  D {\bf 75}, 077701 (2007)
  [arXiv:hep-ph/0611242].

\bibitem{Cheung:2007sva}
  K.~Cheung, J.~Song and Q.~S.~Yan,
  Phys.\ Rev.\ Lett.\  {\bf 99}, 031801 (2007)
  [arXiv:hep-ph/0703149].

\bibitem{Carena:2007jk}
  M.~Carena, T.~Han, G.~Y.~Huang and C.~E.~M.~Wagner,
  JHEP {\bf 0804}, 092 (2008)
  [arXiv:0712.2466 [hep-ph]].

\bibitem{Martin:2007dx}
  A.~Martin,
  arXiv:hep-ph/0703247.


\bibitem{Chang:2006bw}
  S.~Chang, P.~J.~Fox and N.~Weiner,
  Phys.\ Rev.\ Lett.\  {\bf 98}, 111802 (2007)
  [arXiv:hep-ph/0608310].

\bibitem{Forshaw:2007ra}
  J.~R.~Forshaw, J.~F.~Gunion, L.~Hodgkinson, A.~Papaefstathiou and A.~D.~Pilkington,
  JHEP {\bf 0804}, 090 (2008)
  [arXiv:0712.3510 [hep-ph]].

\bibitem{Asai:2004ws}
  S.~Asai {\it et al.},
  Eur.\ Phys.\ J.\  C {\bf 32S2}, 19 (2004)
  [arXiv:hep-ph/0402254].

\bibitem{thomas}
  Work in progress with Thomas Gregoire.

\bibitem{Kaplan:2007ap}
  D.~E.~Kaplan and K.~Rehermann,
  JHEP {\bf 0710}, 056 (2007)
  [arXiv:0705.3426 [hep-ph]].

\bibitem{Gueissaz:2007ds}
  N.~Gueissaz, CERN-THESIS-2007-038.



\end{thebibliography}

\end{document}